\documentclass[reprint,
 amsmath,amssymb,
 aps,
]{revtex4-2}

\usepackage{amsmath,amssymb,dsfont}
\usepackage{xcolor}
\usepackage{graphicx}
\usepackage{hyperref}
\usepackage[normalem]{ulem}
\usepackage{amsfonts}
\usepackage{esint}
\usepackage{float}

\graphicspath{{./}{figures/}}
\newcommand{\bs}[1]{\boldsymbol{#1}}
\newcommand{\dr}[1]{\frac{\mathrm{d}#1}{\mathrm{d}r}}

\definecolor{dark-green}{RGB}{0, 128, 0} 

\definecolor{green2}{RGB}{50, 150, 150} 

\newcommand{\Sect}[1]{\indent{\it {\bf #1}}\\}

\begin{document}

\title{The Exceptional Ring of buoyancy instability in stars}
\author{Armand Leclerc$^1$}
\email{armand.leclerc@ens-lyon.fr}
\author{Lucien Jezequel$^{2}$}
\author{Nicolas Perez$^{1}$}
\author{Asmita Bhandare$^1$}
\author{Guillaume Laibe$^{1,3}$}
\email{guillaume.laibe@ens-lyon.fr}
\author{Pierre Delplace$^{2}$}
\affiliation{
$^1$ Univ Lyon, Univ Lyon1, Ens de Lyon, CNRS, Centre de Recherche Astrophysique de Lyon UMR5574, F-69230, Saint-Genis,-Laval, France.\\
$^2$ Ens de Lyon, CNRS, Laboratoire de physique, F-69342 Lyon, France.\\
$^3$ Institut Universitaire de France.
}

\bibliographystyle{apsrev4-1}
\preprint{APS/123-QED}

\begin{abstract}
We reveal properties of global modes of linear buoyancy instability in stars, characterised by the
celebrated Schwarzschild criterion, using non-Hermitian topology. We identify a ring of Exceptional
Points of order 4 that originates from the pseudo-Hermitian and pseudo-chiral symmetries of the
system. The ring results from the merging of a dipole of degeneracy points in the Hermitian stably-
stratified counterpart of the problem. Its existence is related to spherically symmetric unstable
modes. We obtain the conditions for which convection grows over such radial modes. Those are
met at early stages of low-mass stars formation. We finally show that a topological wave is robust
to the presence of convective regions by reporting the presence of a mode transiting between the
wavebands in the non-Hermitian problem, strengthening their relevance for asteroseismology.
\end{abstract}
\maketitle

\begin{figure*}
    \centering
    \includegraphics[width=\textwidth]{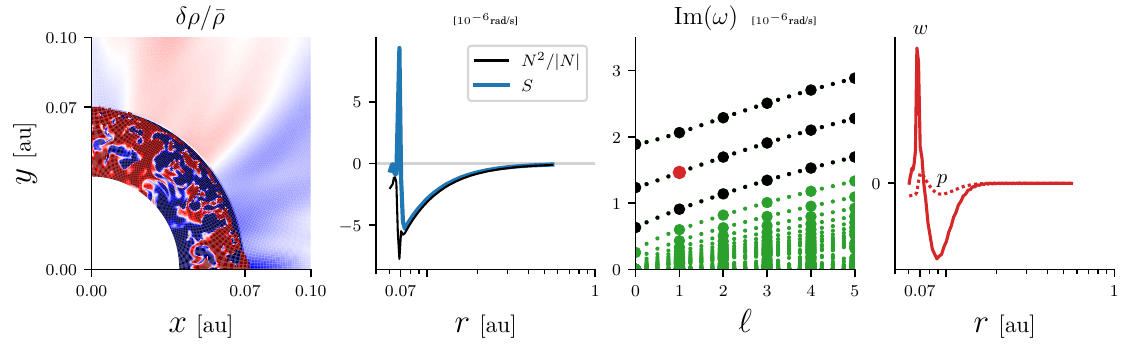}
    \caption{First panel: density fluctuations with respect to azimuthal average, showing convective-like motion. Data from \cite{bhandare2020} (2D simulation of an hydrodynamical stellar collapse), zoomed in near the surface of the protostar. Second panel: average azimuthal profiles of $N^{2}$ and $S$, parameters involved in the linear stability analysis Eq.\eqref{eq:op_pb}. Third and fourth panels: growth rates of a linear stability analysis of this stratification, and profiles of pressure $p$ and radial velocity $w$ of one unstable mode (red), computed numerically with an eigenmodes analysis (see \cite{SM}). The instability develops both inside and outside the surface of the protostar. Three modes (black) differ from the others (green): they have non-zero growth rates on radial perturbations ($\ell=0$). }
    \label{fig:secondCore}
\end{figure*}

A fluid in a gravity field is stratified in density, and results in a stable or an unstable equilibrium. Gravity waves propagate when the stratification is stable, whereas convection develops when the equilibrium is unstable. To develop a convective layer, Sun-like stars must have reached an unstable state where the square of the buoyancy frequency is negative (Schwarzschild criterion $N^2 < 0$, \cite{schwarzschild1906}). Then, through the saturation of a linear instability, the star develops a quasi-adiabatic convective region consisting of large-scale flows that excite waves and transport energy. In these regions, $N^{2}$ takes small negative values for convection to remain sustained, depending on its efficiency (${N^2 \simeq -0.25 \mu\mathrm{Hz}^2}$ in the Sun \citep{kippenhahn1990,lecoanet2013}). Recently, Hermitian topology has shed new light on waves propagating in stably stratified fluids \cite{perrot2019,leclerc2022,perez2022unidirectional}, but the topology of the unstable case, which involves a non-Hermitian formalism, has not yet been studied.
The topological study of waves consists of deducing simple conditions constraining the existence of particular linear modes of physical systems from topological arguments. These arguments can be expressed in a simple way, even for a complicated system of equations. Hermitian systems benefit from general topological index theorems from which one can predict the existence of modes transiting between different wavebands and quantized by a topological integer called the Chern number \cite{Bellissard95,HatsugaiPRL1993,FZ2000,graf2013,Delplace2022}. As such, Hermitian wave topology has become ubiquitous in physical fields as diverse as condensed matter \cite{hasanKane2010}, plasma physics \cite{Parker2020,parker2021,Qin2022}, optics \cite{ozawa2019,lutopological2014}, materials science \cite{xiao2010,Huber2016,Nash14495}, or oceanography \cite{delplace2017,Venaille21, perez2022unidirectional}. 
Recently, topological arguments have been used to reveal the existence of a Lamb-like wave that behaves as a gravity wave at large wavelengths but as a pressure wave at small wavelengths in stably stratified stars \cite{perrot2019,leclerc2022}, raising further questions. Does this wave also propagate in convective regions, which are ubiquitous in stellar objects (e.g. Jupiter or high-mass stars, Fig.~1 of \cite{leclerc2022})? Moreover, seeds of convection in protostars have been observed recently in numerical simulations \citep{bhandare2020,ahmad2023}. Performing a linear stability analysis relative to the background reveals a few unstable radial modes whose origin have not been discussed thus far (see Fig.~\ref{fig:secondCore}). Does topology allows for additional predictions on buoyancy instabilities in stars, to further characterize the physics of the birth of convective layers? To address these questions, we study the non-Hermitian counterpart of the model derived for stellar pulsations. The search for topological properties in non-Hermitian systems has recently stimulated tremendous efforts in condensed matter \cite{delplace2021,jezequel2022a,Ghatak2019}, photonics \cite{zhen2015,Xu2016,Zhang2016}, electric circuits \cite{Kunst18} and geofluids \cite{Zhu2021}, by investigating for instance the existence of topological edge states in non-Hermitian setups, or the appearance of peculiar degeneracy points where the wave operator becomes non-diagonalizable, called exceptional points (EPs). Here, we show that the linear perturbations of a stellar fluid with $N^2<0$ are described by a pseudo-Hermitian and pseudo-chiral symmetric theory. These symmetries constrain the eigenfrequencies, and imply the presence of a ring of EPs of order 4, which is associated with unusual spherically symmetric unstable modes. Furthermore, we report the presence of modes transiting between the complex wavebands of the dispersion relation, one of which is the Lamb-like wave whose topological origin was revealed in \cite{perrot2019}, which we find to be robust to non-Hermitian $N^2<0$ regions.\\

\Sect{Wave operator, wave symbol} We consider a non-magnetic, non-rotating stellar fluid at rest in a spherically symmetric steady state. Perturbations of this equilibrium involve velocity, pressure, and density. Perturbations are adiabatic, modeling stars where the diffusion time is much longer than the dynamical time \cite{schwarzschild1906,ledoux1947}. Perturbations of the gravitational potential are neglected (Cowling approximation \cite{cowling1941}). The equilibrium is still static: no convection has developed yet. We discuss the superadiabatic situation at $N^2 < 0$. We define the perturbation vector $X \equiv  \begin{pmatrix} \Tilde{v} & \Tilde{w} & \Tilde{\Theta} & \Tilde{p} \end{pmatrix}^\top$ based on re-scaled perturbed quantities (respectively horizontal velocity, radial velocity, entropy and pressure), after projection onto vector spherical harmonics of angular number $\ell$ (see SM \cite{SM}). The set of equations for perturbations of the form $e^{-i \omega t} X(r)$ is
\begin{equation}
    \omega X=\mathcal{H} X \ ,
    \label{eq:schro}
\end{equation}
where the \textit{wave operator} $\mathcal{H}$ is defined as
\begin{eqnarray}
&&\mathcal{H}\equiv\label{eq:op_pb}\\
&&\begin{pmatrix}  0 & 0 & 0 & L_\ell(r)\\
0 & 0 & i(N^2)^{1/2} & -iS +\frac{i}{2} c_{\rm s}'+ ic_{\mathrm s}\partial_r\\
 0 & - i(N^2)^{1/2}  & 0 & 0\\
L_\ell(r) & iS +\frac{i}{2} c_{\rm s}'+i c_{\mathrm s}\partial_r & 0 & 0\\ \end{pmatrix}.\nonumber
\end{eqnarray}
This model involves three characteristic frequencies: the squared Brunt-Väisälä frequency
\begin{equation}
    N^2 \equiv -g\dr{\ln\rho_0} - \frac{g^2}{c_\mathrm{s}^2} \ ,
    \label{eq:defN2}
\end{equation}
which characterizes buoyancy, the buoyant-acoustic frequency
\begin{equation}
S \equiv \frac{c_\mathrm{s}}{2g}\left( N^2 - \frac{g^2}{c_\mathrm{s}^2} \right) - \frac{1}{2}\dr{c_{\rm s}} + \frac{c_\mathrm{s}}{r} \ ,
\label{eq:defS}
\end{equation}
which gives the rate at which buoyant and acoustic oscillations exchange momentum \cite{leclerc2022}, and the squared Lamb frequency ${L_\ell^2 \equiv c_\mathrm{s}^2 \; \ell(\ell+1)/r^2}$, which is the momentum in the angular directions. $\rho_0$ is the steady background density, $c_\mathrm{s}$ is the speed of sound and $g$ is the gravity field, which are all functions of the radius $r$. Whenever $N^{2}$ is negative, the fluid is unstable, and the operator $\mathcal{H}$ is non-Hermitian with respect to the standard scalar product.

The spectrum of the model is obtained by solving the system of ordinary differential Eqs.~(\ref{eq:schro},\ref{eq:op_pb}), with appropriate boundary conditions (see SM \cite{SM}). This system implies parameters varying in space, and an analytical solution is in general out of reach. However,  the existence of eigenmodes of $\mathcal{H}$ such as Lamb-like modes, whose frequency transits between other modes when varying a parameter (here $\ell$), can be easily accessed without explicitly solving the differential system, but through topological properties of a dual \textit{wave symbol}, a matrix $H$ with scalar coefficients obtained by a Wigner transform of the wave operator $\mathcal{H}$ that maps  the differential problem onto phase space \cite{Delplace2022}, as suggested by \cite{onuki2020}.
$H$ represents physically the local action of the medium on a plane wave, without requiring that the medium varies slowly with respect to the wavelength (see SM \cite{SM}). This symbol matrix $H$ reads 
\begin{equation}
H \equiv
\begin{pmatrix}
    0 & 0 & 0 & L_\ell \\
    0 & 0 & iN & K_r - iS \\
    0 & -iN & 0 & 0 \\
    L_\ell & K_r +iS & 0 & 0
\end{pmatrix} ,
\label{eq:symbol}
\end{equation}
and depends on the 3 parameters $K_r$, $L_\ell$ and $S$ for fixed $N^2$. $K_r = c_\mathrm{s}k_r$ with $k_r$ the Wigner symbol of $-i\partial_r$ is the radial wavenumber of a wave locally plane. We denote $\omega$ and $\Omega$ the eigenvalues of $\mathcal{H}$ and $H$ respectively.
When $N^{2} > 0$, the matrix $H$ is Hermitian and always diagonalizable with real eigenvalues. When $N^2<0$, $N$ is purely imaginary and $H \neq \bar{H}^\top$.\\

\Sect{Symmetries and Exceptional Points}
For a subset of the parameter space $\left( K_r,S, L_\ell\right)$, $H$ is non-diagonalizable. These particular points are EPs. At these points, the eigenvalues are degenerate and the eigenvectors coalesce, in the sense that the number of independent eigenvectors is less than the number of eigenvalues that merge. The occurrence of EPs is constrained by the presence of certain symmetries. In our case, one notices that $H$ benefits from a pseudo-Hermitian symmetry
\begin{align}
UHU^{-1} = \Bar{H}^\top   ,
\label{eq:psH}
\end{align}
with the unitary transform $U=\text{diag}(1,1,-1,1)$. Eigenvalues of pseudo-Hermitian matrices are either real or complex conjugate pairs. Pseudo-Hermiticity also increases the order of EPs in the parameter space \cite{delplace2021}. $H$ also has a chiral symmetry ${\Gamma H \Gamma^{-1} = -H}$, with the unitary transform $\Gamma=\text{diag}(1,1,-1,-1)$, which can be traced back from the time-reversal symmetry of the fluid lagrangian. Equivalently, this chiral symmetry combined with the pseudo-Hermitian symmetry (Eq.~\ref{eq:psH}) can be taken into account as a pseudo-chiral symmetry 
\begin{align}
(\Gamma U) H (\Gamma U)^{-1} = -\bar{H}^\top ,
\label{eq:psChiral}
\end{align}
that was also shown to constrain the existence of EPs \cite{delplace2021}. We show that the combined effect of both pseudo-chirality and pseudo-Hermiticity leads to a codimension 2 for 4-fold EPs (see SM \cite{SM}). This means that, for ${N^2<0}$, the 4 complex-valued eigenbands of $H$ are expected to cross on a curve in the ($K_r,S,L_\ell$) space. A direct derivation shows that those EPs satisfy
\begin{eqnarray}
    L_\ell&=&0 , \\  K_r^2 + S^2 &=& - N^2 ,
\label{eq:condEP}
\end{eqnarray}
meaning that they form a circle of radius $|N|$ around the origin in the $(K_r,S)$ plane at $L_\ell=0$. $H$ is diagonalizable everywhere apart from this circle, where only two eigenvectors exist, $\begin{pmatrix} 1 & 0 & 0 & 0 \end{pmatrix}^\top$ and $\begin{pmatrix} 0 & 0 & (i K_r+S)/N & 1 \end{pmatrix}^\top$. This \textit{exceptional ring} thus consists of 4-fold EPs (algebraic multiplicity of $4$) with a geometric multiplicity of $2$.

\begin{figure}
    \centering
    \includegraphics[width=\columnwidth]{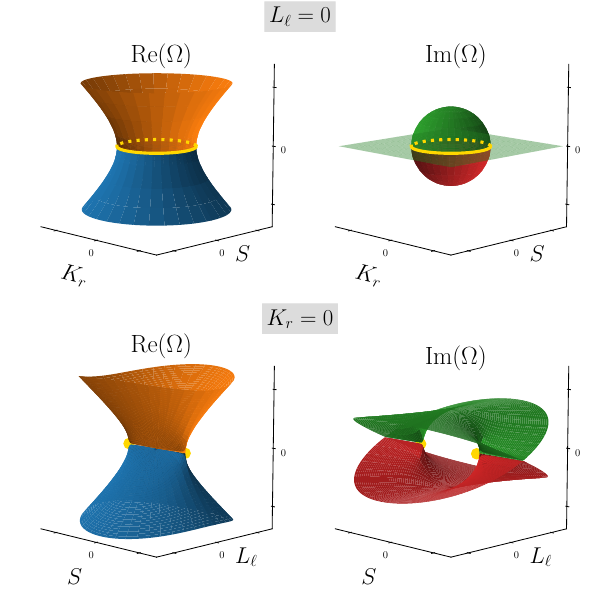}
    \caption{Eigenvalues $\Omega$ of $H$ around the EPs. There are two acoustic bands (orange and blue) with real eigenvalues, and two gravity bands (green and red) with purely imaginary eigenvalues. The yellow rings and points highlight the positions of the Exceptional Points. At large wavelengths $K_r \lesssim N$, pulsations and onset of convection behave very differently from what is expected in the short wavelength limit or in a Boussinesq approximation. Top right: ``bubble of instability'' \cite{mackay2020stability}. Bottom left: ``double-coffee-filter'' \cite{kirillov2021nonconservative}. Bottom-right: ``viaduct'' \cite{kirillov2021nonconservative}.}
    \label{fig:freqs_ep3s}
\end{figure}

This ring where modes degenerate separates radial modes ($\ell=0$) into two regions of distinct spectral properties. Figure~\ref{fig:freqs_ep3s} shows the real and imaginary parts of the eigenvalues of $H$. 
Outside the ring ($K_r ^{2} + S^{2} > \vert N^{2} \vert$), the radial modes behave classically \cite{tassoul1967}: radial pressure waves have finite real frequencies and radial buoyancy modes have zero growth rates. Inside ($K_r ^{2} + S^{2} < \vert N^{2} \vert$), they behave differently: the acoustic bands degenerate at $\Omega = 0$, and gravity modes have non-zero growth rates, the maximum value $\sqrt{-N^2}$ being reached for $K_r = S = 0$. When crossing the ring, two eigenvalues of $H$ transit from real to pure imaginary values. Since $H$ is pseudo-Hermitian, this can be interpreted as a Krein collision in the framework of Krein signature theory \cite{kirillov2021nonconservative}. Unstable (imaginary) eigenvalues with zero Krein signature unfold from the encounter of stable (real) eigenvalues with opposite Krein quantities $\kappa \left( X \right) = \bar{X}^\top U X$, $X$ being the corresponding eigenvector of $H$, colliding at the EP ring. A Krein quantity $\tilde{\kappa} = \int \mathrm{d}r\mathrm{d}\Phi\; (|\tilde{v}|^2 + |\tilde{w}|^2 +|\tilde{p}|^2 -|\tilde{\Theta}|^2)$, with $\Phi$ the solid angle, can also be defined for any solution $X(\bs{r},t)$ of Eq.\eqref{eq:schro} and is a conserved quantity of the flow. In particular, $\tilde{\kappa} =0$ for an unstable mode (see SM \cite{SM}).

\begin{figure}
    \centering
    \includegraphics[width=\columnwidth]{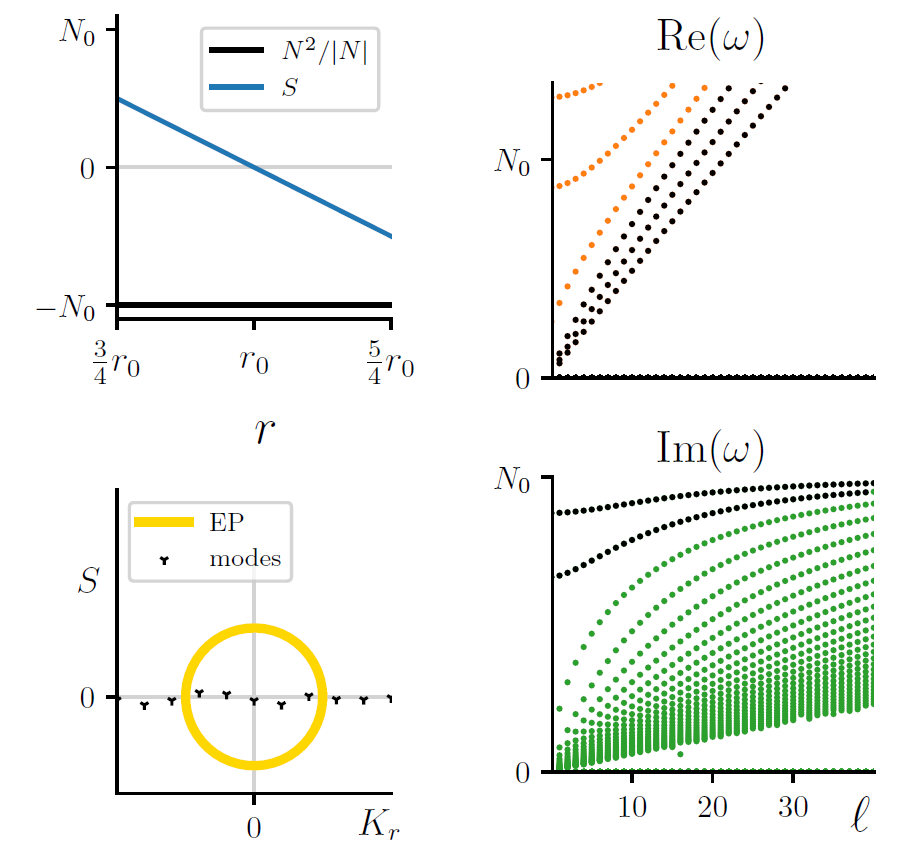}
    \caption{Spectrum of a model with the stratification profile shown in the top left panel, corresponding to a layer $N^{2} < 0$ such that $S$ goes to zero at a given radius $r_0$ and remains smaller than $\vert N \vert$. Right panels: i) three acoustic modes with $\omega(\ell=0)=0$ and ii) two gravity modes have non-zero growth rates in the $\ell = 0$ limit. The other modes have a classical behavior. Bottom left: schematic of the location of the modes in parameter space: 5 modes behave differently because they are inside the EP ring. Orange points are acoustic modes, green points are unstable buoyancy modes, black points are Exceptional modes. See SM for details on numerics \citep{SM}.}
    \label{fig:pinching}
\end{figure}

To date, no theorem connects the EPs of the symbol matrix $H$ to a possible manifestation in the spectrum of $\mathcal{H}$. If such a connection exists, one expects to find the footprint of EPs in radial modes ($\ell = 0$) as this is where the EP ring is found in the Wigner matrix, when the radial wavelength is large enough and the profiles of $N^{2}$ and $S$ are such that the parameters cross the ring shown in Fig.~\ref{fig:freqs_ep3s} as $r$ varies. Furthermore, the above analysis suggests that the relevant unstable modes are those of wavelengths typically longer than $\sim c_s /\vert N\vert$ ($N^2 \neq 0$ since convection has not started nor saturated to a quasi-adiabatic state yet). This condition also requires $S$ to be smaller than $\vert N\vert$, at least locally. Figure~\ref{fig:pinching} shows the spectrum of a model where the aforementioned condition is satisfied. The unstable region is wide enough so that low-order radial modes have a sufficiently large radial wavelength, enough for the corresponding $K_r$ to be located inside the ring. The spectrum exhibits three acoustic waves with zero frequency for $\ell \rightarrow 0$ and two unstable buoyancy modes with non-zero growth rates for $\ell \rightarrow 0$. Various profiles of pre-convective unstable equilibria have been tested (Fig.4 of SM \cite{SM}). They all have such exceptional modes since they are continuous deformations of the model of Fig.~\ref{fig:pinching}. Additional modes enter the EP ring by pairs when increasing the length of the layer. These properties are a physical footprint of the existence of EPs. These results are consistent with recent reports of experiments on compressible fluids, in which convection develops via axisymmetric modes \cite{menaut2019,koulakis2021,koulakis2023}.\\

\Sect{Fundamental mode} In the stably stratified problem ($N^2>0$), $H$ has degenerated eigenvalues for $(K_r,S,L_\ell)=(0,0,\pm N)$ for which both the gravity and acoustic waves have frequencies $N$. Such degeneracies act as monopoles of Berry curvature in the parameter space $(K_r,S, L_\ell)$, and carry  topological charges given by Chern numbers $\pm 1$. Those Chern numbers are in direct correspondence with the existence of the Lamb-like waves in the spectrum of the operator $\mathcal{H}$, and explain the transit of the fundamental mode between the bands \cite{perrot2019, leclerc2022, perez2022unidirectional}.
\begin{figure}
    \centering
    \hspace*{-0.5cm}\includegraphics[width=1.1\columnwidth]{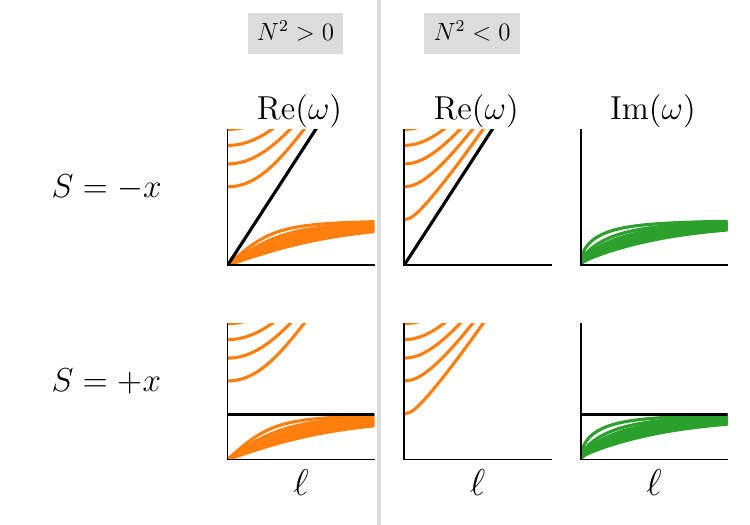}
    \caption{Frequencies of models with $S$ varying linearly in space (on some appropriately rescaled spatial variable $x$). Left: Stable stratification. The transiting mode depends on the sign of $\mathrm{d}S/\mathrm{d}x$. Right: Unstable stratification. Apart from the buoyancy modes being transposed to imaginary values, the transiting mode behaves as it does in the stable case. When $S = -x$, it arises as a propagating Lamb-like wave. When $S=+x$, it is an unstable mode of growth rate $\vert N\vert$, independently of $\ell$. Only the first 10 modes of each band are represented.}
    \label{fig:NHspecFlow}
\end{figure}

In the present study, $H$ is no longer Hermitian, and the correspondence between the Lamb-like wave and the Chern numbers is not guaranteed.
Several approaches have recently been developed to address the topological properties of non-Hermitian operators \cite{Gong2018,Ghatak2019,Deng19,Yao2018,Borgnia20,Kunst18,NonHermiPhysics,Kunst2021,TopoBandNonHerm,NonHermiPhysics, TonyLee16,Xu2016,Zhang2016}. In particular, non-Hermitian formulations of the Chern numbers as monopoles of Berry curvature have been proposed, and a non-Hermitian generalization of the correspondence with the transit of the fundamental mode has been developed \cite{WeylChern,jezequel2022a}. However, such a generalization cannot apply here, as the Hermitian degeneracy point is turned into a EP curve when the sign of $N^2$ is swapped, with zero net Chern number. Other works have introduced winding numbers associated to such circles of EPs \cite{Xu2016,Zhang2016,WeylChern}, which we also find to vanish here. Nevertheless, we confirm below the existence of the the Lamb-like wave in regions with $N^2<0$. To do so, we study the normal form, setting linear spatial dependency for $S$, that is $ S\left(r \right) = \alpha \left(r - r_{0} \right)$, and $N^2 < 0$, sound speed $c_s$ and Lamb frequency $L_\ell$ constant \cite{venaille2023,leclerc2022}. The spectral properties of this problem capture the essential topology that will be reflected in the spectra of real objects. Within these assumptions, Eq.~\eqref{eq:schro} is found to admit a fundamental mode with zero node trapped around the radius $r_0$ where $S(r_0)=0$. However, its behaviour depends strongly on the slope of $S$ at $r_0$, as shown in Fig.~\ref{fig:NHspecFlow} (derivation in SM \cite{SM}). For a negative slope ($\alpha<0$) this mode verifies $\omega^2 = L_\ell^2$ and its eigenfunctions are $\Tilde{v},\Tilde{p} \; \propto \exp\left(-\frac{\alpha}{2c_\mathrm{s}}(r-r_0)^2\right)$, $\Tilde{w}=\Tilde{\Theta}  =0$, which have the peculiar property of having no radial velocity nor entropy perturbation.
This is the Lamb-like wave, and we thus conclude that it still propagates for $N^2<0$. In contrast, for a positive slope ($\alpha>0$), the fundamental mode verifies $\omega^2 =-{\vert N^2 \vert}$ and corresponds to a growing perturbation. Its eigenfunctions are $\Tilde{v}=\Tilde{p} = 0$, $\Tilde{w},\Tilde{\Theta} \; \propto \exp\left(-\frac{\alpha}{2c_\mathrm{s}}(r-r_0)^2\right)$, which have no angular velocity or pressure perturbation. We verified numerically that this mode is independent of the boundary conditions (see SM \cite{SM}). The importance of polarization relations is key for wave topology \cite{onuki2020,perez2021manifestation,venaille2022}. Equation~\eqref{eq:schro} admits non-zero solutions even if some of the component fields are equal to zero. Preserving the vector structure of the problem prevents the filtration of such solutions, as it may happen when decoupling the initial system of equations into a single high-order ordinary differential equation. The general problem is expected to have the same properties, since it is a continuous deformation of this model, as long as no new location where $S$ goes to zero is introduced (Fig.4 of SM \cite{SM}). When $N^2(r)$ takes positive and negative values in different regions of the star, the Lamb wave still exists and coexists with an unstable buoyancy band. This is true whether $S(r)$ goes to zero inside the stable or unstable region. In sharp contrast, when the profile of $S(r)$ goes to zero with a positive slope in a region of negative $N^2$, we observe an unstable mode with a growth rate $\sim\sqrt{\vert N^2 \vert}$, independently of $\ell$.\\

\Sect{Asteroseismology}
The topological study of pulsating modes in stars has so far been restrained to radiative regions ($N^{2} > 0$), the problem being Hermitian \cite{leclerc2022}. The question of whether the Lamb-type topological wave could propagate in convective regions (small $N^{2} < 0$) remained unanswered. We show in this study that these waves can indeed propagate within them. They are therefore relevant even for objects such as high mass stars or Jupiter (see Fig.~1 of \cite{leclerc2022}). On top of this, convective regions can also generate multiple exceptional modes that behave like acoustic waves with zero frequency at $\ell = 0$. The existence or not of such modes in observational data constrains the internal structure of objects with convective interiors.\\

\Sect{Birth of convection in protostars}
Unstable exceptional modes of low radial order, low $\ell$ and high growth rates develop when the conditions $N^{2}< 0$ and $N^2 + S^2 <0$ are satisfied. These conditions are met during the formation of a low-mass protostar, as shown in Fig.~\ref{fig:secondCore} from 2D simulations \cite{bhandare2020} (see SM \cite{SM} for physical interpretation). This clarifies the origin of radial unstable modes developing around the surface of the protostar. Hence, topological modes provide a possible explanation for the long-lasting problem of how and when convection starts in young stars. Further high-resolution 3D numerical simulations are however required to prove that the kinematic signature observed correspond indeed to convective motion, and to study how these modes will develop in the non-linear regime (e.g. convective eddies or fully developed turbulence).\\

Future studies are needed to quantify the role of rotation and self-gravity on these modes. Additional symmetries are expected to be broken in some regions of the extended parameter space. Exceptional Points and Krein signature will be key tools to diagnose properties of global modes in such complex objects. The topological invariant associated with exceptional modes remains to be found.\\

\begin{acknowledgments}
We acknowledge funding from the ERC CoG project PODCAST No 864965. PD is supported by the national grant ANR-18-CE30-0002-01. AL and LJ are funded by a Contrat Doctoral Spécifique Normaliens. We thank A.Marie, G. Chabrier, E. Lynch, M. Rieutord, F. Lignières, B. Commerçon, I . Baraffe and A. Le Saux for useful comments and discussions.
\end{acknowledgments}
\bibliographystyle{unsrtnat}
\typeout{}

\clearpage
\appendix
\begin{widetext}
\section{Rescaled quantities}
\label{app:rescale}
Equations of mass, momentum and energy conservation are linearized around a spherically symmetric steady state. Denoting $\rho_0$, $c_\mathrm{s}$ and $g$ the density, sound speed and gravitational acceleration of the equilibrium, and rescaling the perturbation quantities by
\begin{equation}
  \begin{aligned}
    \bs{v}' &\mapsto  \Tilde{\bs{v}} = \rho_0^{1/2}r \;\bs{v}',\\        
    p' &\mapsto \Tilde{p} = \rho_0^{-1/2}c_\mathrm{s}^{-1}r\;p',\\
\rho' &\mapsto \Tilde{\Theta} = \rho_0^{-1/2}r\frac{g}{(N^2)^{1/2}}\;(\rho' - \frac{1}{c_\mathrm{s}^2}p'),
  \label{eq:transform}
  \end{aligned}
\end{equation}
one obtains the multi-component equation ${i\partial_t X = \mathcal{H}_5 X}$ for rescaled perturbed quantities, where

\begin{eqnarray}
\mathcal{H}_5 &\equiv& 
\begin{pmatrix} 0 & 0 & 0 & 0 & -i\frac{c_{\rm s}}{r\sin(\theta)}\partial_\phi\\
0 & 0 & 0 & 0 & -i\frac{c_{\rm s}}{r}\partial_\theta\\
0 & 0 & 0 & i(N^2)^{1/2}\;\;\;\; & -iS +\frac{i}{2} c_{\rm s}'+ ic_{\mathrm s}\partial_r\\
0 & 0 & - i(N^2)^{1/2}  & 0 & 0\\
-i\frac{c_{\rm s}}{r\sin(\theta)}\partial_\phi\;\;\;\; & -i\frac{c_{\rm s}}{r\sin(\theta)}\partial_\theta({\scriptstyle \sin(\theta)} \cdot)\;\;\;\; & iS +\frac{i}{2} c_{\rm s}'+i c_{\mathrm s}\partial_r & 0 & 0\\ \end{pmatrix},\label{eqApp:op_pb}\\
\nonumber\\
X &\equiv& \begin{pmatrix}\Tilde{v_\phi} & \quad\Tilde{v_\theta} & \quad\Tilde{w} & \quad\Tilde{\Theta} & \quad\Tilde{p}\end{pmatrix}^\top.
\end{eqnarray}

\section{Vector spherical harmonics}
\label{app:VSH}
We project the perturbations on vectorial spherical harmonics
\begin{eqnarray}
    \bs{Y}_\ell^m &\equiv& Y_\ell^m \bs{e}_r,\\
    \bs{\Psi}_\ell^m &\equiv& \frac{ir}{\sqrt{\ell(\ell+1)}} \bs{\nabla}Y_\ell^m \quad\quad\;\;\: \text{or 0 if }\ell=0,\\
    \bs{T}_\ell^m &\equiv& \frac{i}{\sqrt{\ell(\ell+1)}}\bs{r}\wedge\bs{\nabla}{Y}_\ell^m \quad\text{or 0 if }\ell=0.
\end{eqnarray}
$Y_\ell^m$ is the spherical harmonic function of harmonic and azimuthal degrees $(\ell,m)$, $\bs{e}_r$ is the unit vector in the radial direction. The normalization slightly differs from \cite{barrera1985}. For $\ell > 0$, those functions are orthonormal with respect to the scalar product on the sphere
\begin{eqnarray}
    \int\mathrm{d}\Phi\quad \bs{Y}_\ell^m\cdot \bs{Y}_{\ell'}^{m'*} &=& \delta_{\ell\ell'}\delta_{mm'},\\
    \int\mathrm{d}\Phi\quad \bs{\Psi}_\ell^m\cdot \bs{\Psi}_{\ell'}^{m'*} &=& \delta_{\ell\ell'}\delta_{mm'},\\
    \int\mathrm{d}\Phi\quad \bs{T}_\ell^m\cdot \bs{T}_{\ell'}^{m'*} &=& \delta_{\ell\ell'}\delta_{mm'},\\
    \int\mathrm{d}\Phi\quad \bs{Y}_\ell^m\cdot \bs{\Psi}_{\ell'}^{m'*} &=& \int\mathrm{d}\Phi\quad \bs{\Psi}_\ell^m\cdot \bs{T}_{\ell'}^{m'*}
    = \int\mathrm{d}\Phi\quad \bs{T}_\ell^m\cdot \bs{Y}_{\ell'}^{m'*} =0.
\end{eqnarray}
$\Phi$ is the solid angle. The rescaled perturbed velocity field can be decomposed as $\Tilde{\bs{v}} = \Tilde{w}(r)\bs{Y}_\ell^m + \Tilde{v}(r)\bs{\Psi}_\ell^m$, along with ${\Tilde{\rho},\Tilde{p} \propto Y_\ell^m}$. Using relations on vectorial spherical harmonics, one has 
\begin{eqnarray}
    \partial_t \Tilde{v} &=& -\int\mathrm{d}\Phi \quad \bs{\nabla}(\Tilde{p}')\cdot {\bs{\Psi}_\ell^m}^* = \frac{\sqrt{\ell(\ell+1)}}{r}\Tilde{p}' = \frac{\sqrt{\ell(\ell+1)}}{r}\Tilde{p}',\nonumber\\
    \\
    \partial_t \Tilde{p} &=& \int\mathrm{d}\Phi\quad Y_\ell^{m*} \left( (S+\frac{c_\mathrm{s}'}{2}+c_\mathrm{s}\partial_r)\Tilde{w}Y_\ell^m+\mathrm{div}(\Tilde{v}\bs{\Psi_\ell^m}) \right)
    = (S+\frac{c_\mathrm{s}'}{2}+c_\mathrm{s}\partial_r)\Tilde{w} -\frac{\sqrt{\ell(\ell+1)}}{r}\Tilde{v}.
\end{eqnarray}
Introducing the Lamb frequency $L_\ell^2 = c_\mathrm{s}^2\frac{\ell(\ell+1)}{r^2}$, one obtains the 4x4 system of equations Eq.~\eqref{eq:op_pb} defining $\mathcal{H}$.

\section{Numerical calculation of the eigenmodes}
\label{app:dedalus}
The problem Eq.\ref{eq:op_pb}, along with the boundary conditions $\Tilde{w}=0$ on both sides of the domain, is an eigenvalue problem of the differential operator $\mathcal{H}$. We used the \textsc{EVP} problem class of the python package \textsc{Dedalus} \cite{burns2020} to numerically solve this problem. \textsc{Dedalus} uses spectral methods, and decomposes solutions on $N_r$ Chebyshev polynomials, in order to obtain a matrix eigenvalue problem which is solved by linear algebra techniques. The spatial resolution is given by $L/N_r$, with $L$ the domain size. At most $4 N_r$ eigenmodes are possible to find for a given resolution. To ensure numerical convergence on the eigenmodes, we used the \textsc{eigentools} package to reject spurious or unresolved modes \cite{oishi2021}. Each problem is solved twice, first with resolution $N_r$ then with resolution $1.5\,N_r$. If the eigenvalue changed significantly, the mode is unresolved and is rejected. Quantitatively, the user chooses a threshold $\delta$ such that if $\vert \omega_{n,1.5 N_r} - \omega_{n,N_r} \vert/\vert\omega_{n,N_r} \vert > \delta$, the mode is rejected. \\
A given model takes as an input the functions $S(r)$, $N(r)$ and $c_\mathrm{s}(r)$, and $\ell$ is a parameter. For each value of $\ell$, we solve for $N_r$ eigenmodes and store their profiles and complex eigenfrequencies $\omega$, in order to obtain the points $\omega(\ell)$ presented on panel 3 of Fig.\ref{fig:secondCore} and right panels of Fig.\ref{fig:pinching}. The finite resolution explains why modes with low $\mathrm{Im}(\omega)$ are not fully determined: they are modes with high radial order, unresolved at the current resolution. For the model with linear spatial dependency of $S$ on Fig.\ref{fig:pinching}, we used $N_r = 64$ polynomials and $\delta = 10^{-8}$. The python script used can be found on \url{https://github.com/ArmandLeclerc/ExcepRing_convection}. For the stability analysis of the protostar on Fig.\ref{fig:secondCore}, we used $N_r = 256$ polynomials, in order to resolve the sharp features of the model, and $\delta = 10^{-3}$.

\end{widetext}
\clearpage

\section{Outer boundary condition}
\label{app:BCs}
\begin{figure}
    \centering
    \includegraphics[width=0.85\columnwidth]{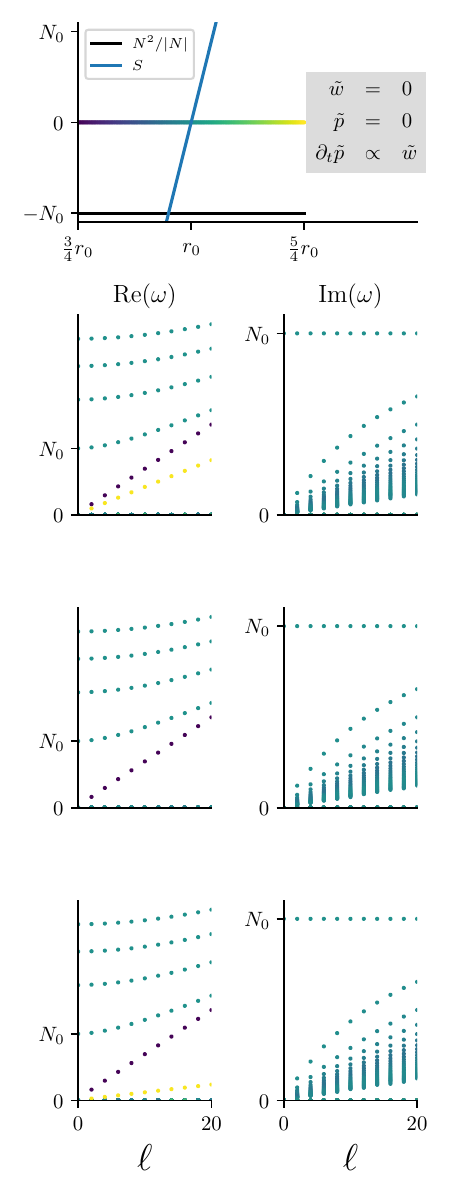}
    \caption{Influence of boundary conditions. Top: model solved for different outer boundary conditions listed on the right. The sound speed is assumed to be constant. The inner boundary condition is always $\Tilde{w}=0$. Next three rows: The three associated spectra computed numerically, in the same order as listed in the top panel. They show the real band (left) and the imaginary band (right), the colors of the modes show their average localisation $\langle r\rangle$. Therefore blue modes are bulk modes, and boundary modes are in dark purple/ light yellow. Only the outer boundary mode in light yellow changes when changing the boundary condition, none of the other change in frequency significantly.}
    \label{fig:NF_BC}
\end{figure}

The outer boundary condition needs caution to ensure treatment of potential singularity at surface as ${\rho_0=0}$~\cite{gough1993}. The modes of interest are bulk modes, and do not change significantly as long as they are localized far away from the surface and the boundary condition imposed. Numerically, imposing $\Tilde{w}(R)=0$, $\Tilde{p}(R) = 0$ or a free-surface-like condition $\partial_t\Tilde{p} \propto \Tilde{w}$ only changes the existence of boundary modes, and has no impact on the modes of interest described in this study (see Fig.~\ref{fig:NF_BC}). 

\clearpage

\begin{widetext}
\section{Wigner transform}
\label{app:matrix}
We use the Wigner transform as defined in Appendix C of \cite{leclerc2022}. It transforms the differential operator $\mathcal{H}$ (Eq.\eqref{eq:op_pb}) into a matrix of scalars, providing a representation in a phase space $\{r,k_r\}$. Elementary Wigner transforms are
\begin{eqnarray}
    f(r) &\mapsto& f(r) \text{ for any function }f,\\
    i\partial_r &\mapsto& k_r,\\
    ic_\mathrm{s}\partial_r &\mapsto& c_\mathrm{s}k_r - i\frac{c_\mathrm{s}'}{2}.
\end{eqnarray}
Applying these identities to $\mathcal{H}$, one obtains the Wigner Symbol that reads
\begin{equation}
H =
\begin{pmatrix}  0 & 0 & 0 & L_\ell\\
0 & 0 & i(N^2)^{1/2}\;\;\;\; & -iS +K_r\\
 0 & - i(N^2)^{1/2}  & 0 & 0\\
L_\ell\;\;\; & iS + K_r & 0 & 0\\ \end{pmatrix},
\label{eqApp:4x4Symb}
\end{equation}
where 
\begin{equation}
    K_r \equiv c_\mathrm{s}k_r,
\end{equation} where $k_r$ is the Wigner symbol of $i\partial_r$, representing the radial wavelength of the wave. The Wigner transform was introduced in quantum physics \cite{weyl1927quantenmechanik,wigner1932quantum} and has later been used in other fields to capture geometric corrections in the ray-tracing dynamics of multi-component classical waves in fluid and elastic media \cite{littlejohn1991geometric,emmrich1996geometry,ryzhik1996transport,vanneste1999wave,perez2021manifestation,venaille2023}. The Wigner transform provides a rigorous and convenient way to perform a local analysis of a differential operator \cite{onuki2020}. It maps differential operators and functions on a higher-dimensional space, the phase space $(r,k_r)$, that are treated as independent variables. It does not assume scale separation between wavelengths and typical lengths over which the background varies, and as such it differs from a JWKB approximation \cite{onuki2020}.

\section{Characterization of Exceptional Points}
\label{app:EPs}

\begin{figure}[H]
    \centering
    \includegraphics[width=0.4\textwidth]{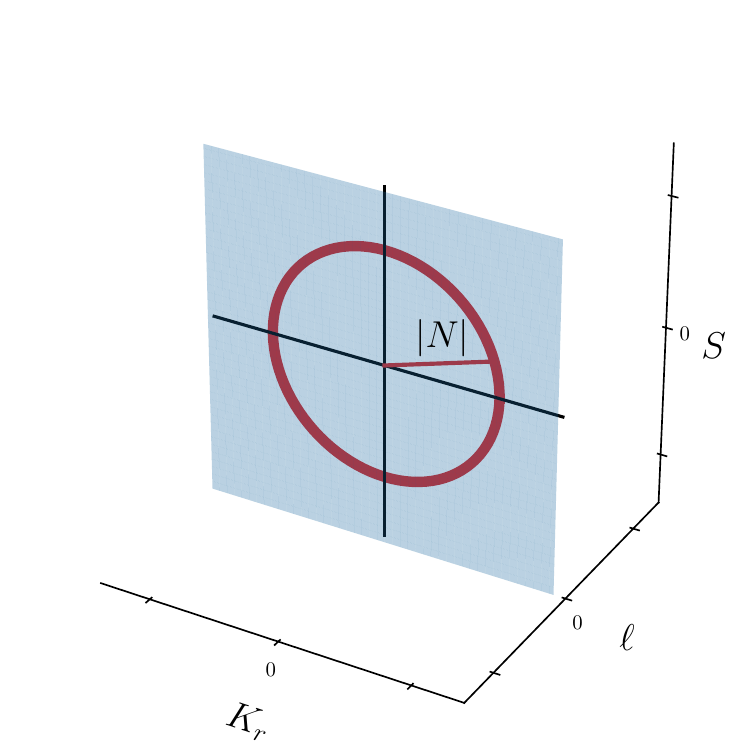}
    \caption{Localizations of degeneracies and Exceptional Points of $H$ in the parameter space. The blue plane at {$\ell = 0$} is a double degeneracy plane, since outside the red circle $H$ is diagonalizable and its eigenvalue 0 is of multiplicity 2. The red circle corresponds to a coalescence of three eigenvectors. These points are Exceptional Points of order 4, where the matrix $H$ is no longer diagonalizable.}
    \label{fig:EP3}
\end{figure}

Exceptional Points are identified by applying the procedure described in \cite{delplace2021} to the symbol matrix $H$. 
Isolated $4$-fold EPs (requiring the crossing points of $4$ complex-valued bands) are expected to appear in three-dimensional parameter space provided either pseudo-Hermitian symmetry or pseudo-chiral symmetry applies. This property directly follows from the characteristic polynomial $P(x)=\det(H-x \mathds{1})$, that expands as
\begin{align}
 P(x)=a_0 + a_1 x + a_2 x^2 + a_3 x^3 + a_4 x^4 + \dots   
\end{align}
and has real coefficients $a_j$ in the pseudo-Hermitian case, while in the pseudo-chiral case only the even terms $a_{2j}$ are real, the odd terms $a_{2j+1}$ being purely imaginary. Therefore, in our case, where the symbol matrix $H$ is both pseudo-Hermitian and pseudo-chiral symmetric, all odd coefficients vanish and we are left with 
$P(x) = a_0 +  a_2 x^2  + a_4 x^4$ (and we take $a_4=1$ without loss of generality). Looking for 4-fold EPs thus amounts to looking for the conditions for which the resultants $ R_1 \equiv \mathcal{R}(P,P')$, $R_2 \equiv \mathcal{R}(P',P'')$ and $R_3 \equiv \mathcal{R}(P'',P''')$ vanish, where $\mathcal{R}(Q_1,Q_2)$ is the resultant of two polynomials $Q_1$ and $Q_2$. Since with either pseudo-Hermitity or pseudo-chirality, those resultants are purely real or purely imaginary \cite{delplace2021}, this yields $3$ constraints to satisfy. However, in the presence of both symmetries, the number of constraints is reduced to $2$ as we can see from the general expressions
\begin{align}
 &R_1 = a_0 (16 a_0 - 4 a_2^2)^2\\
 &R_2 = (144 a_0 - 20 a_2^2)^2 \\
&R_3 = 331776 a_0 
\end{align} 
since the vanishing of $R_3$ automatically induces also the vanishing of $R_1$. The condition $a_0=a_2=0$ cancels all three and represents EP4s, which we express on the parameters of $H$ below.\\
The codimension of 4-fold EPs in the presence of both pseudo-Hermiticity and pseudo-chirality is thus $2$, so we expect to find a line of such EPs in three-dimensional parameter space. In the specific case of our symbol matrix $H$, one  finds
\begin{eqnarray}
    R_1 
    &=& -4 L_\ell^2 N^2 \left( (K_r^2 + S^2 + L_\ell^2 +N^2)^2 - 4L_\ell^2N^2\right)^2,\\
    R_2 
    &=& -512 \left(K_r^2 + L_\ell^2 + S^2 + N^2  \right)^3,\\
    R_3 
    &=& -1152\left(K_r^2 + L_\ell^2 + S^2 + N^2  \right).
\end{eqnarray}
$R_1$ cancels only for radial modes $\ell = 0$. For such modes, $R_{2}$ and $R_{3}$ cancel when the following condition is satisfied
\begin{equation}
    L_\ell=0 \: \text{ and } \: K_r^2 + S^2 + N^2 = 0.
\end{equation}
Since $N^2 < 0$, Eq.~\eqref{eq:condEP} is that of a circle in space $\{K_r,S\}$, centered on the origin and of radius $\vert N\vert$. On this circle, the four eigenvalues of $H$ are equal ($\Omega_{1,2,3,4} = 0$). When $L_\ell = 0$, $H$ generically has 4 eigenvectors:
\begin{eqnarray} \label{eq:eigenvectorss}
    &&\begin{pmatrix}
        1\\0\\0\\0
    \end{pmatrix},
    \begin{pmatrix}
        0\\0\\ (i K_r + S)/{N} \\1
    \end{pmatrix},
    \begin{pmatrix}
        0\\ \sqrt{K_r^2 + S^2 + N^2}/{(i K_r + S)} \\ {N}/{(i K_r - S)} \\1
    \end{pmatrix},
    \begin{pmatrix}
        0\\ -{\sqrt{K_r^2 + S^2 + N^2}}/{(i K_r + S)} \\ {N}/{(i K_r - S)} \\1
    \end{pmatrix} .
\end{eqnarray}
Three cases can then be distinguished. When $L_\ell \neq 0$, $H$ has four distinct eigenvalues and four distinct eigenvectors ($R_{1} \neq 0$). When $L_\ell = 0$ and $K_r^2 + S^2 + N^2 \neq 0$, $H$ has three distinct eigenvalues but four distinct eigenvectors ($R_{1} = 0$, $R_{2,3} \neq 0$). Such a region where the eigenvalues degenerate but where the eigenvectors do not coincide is infrequent in non-Hermitian systems \cite{delplace2021}. When $\ell = 0$ and $K_r^2 + S^2 + N^2 = 0$ ($R_{1,2,3}=0$), $H$ has one degenerate eigenvalue and the only two distinct eigenvectors
\begin{equation}
    \begin{pmatrix}
        1\\0\\0\\0
    \end{pmatrix},
    \begin{pmatrix}
        0\\0\\ (i K_r + S)/{N}  \\1
    \end{pmatrix}.
\end{equation}
The parameter space 
can be endowed with cylindrical coordinates, with ${K_r=\rho\cos(\psi), S = \rho\sin(\psi)}$. In these coordinates, the two eigenvectors read
\begin{equation}
    \begin{pmatrix}
        1\\0\\0\\0
    \end{pmatrix},
    \begin{pmatrix}
        0\\0\\ i\mathrm{e}^{-i\psi}  \\1
    \end{pmatrix}.
\end{equation}

\section{Krein signatures and collisions}
\label{app:krein}

Let $X$ be an eigenvector of $H$ of eigenvalue $\Omega$. Define the Krein quantity $\kappa\left( X\right)$ associated to the  unitary diagonal matrix $U = \text{diag}(1,1,-1,1)$ of Eq.~\eqref{eq:psH} as
\begin{equation} \label{eq:Krein_signature}
    \kappa(X) \equiv \bar{X}^\top U X ,  
\end{equation}
and $\sigma =\mathrm{sgn} \,\kappa$ its associated Krein signature. The equality $\kappa (X) = 0$ holds whenever $X$ corresponds to an unstable eigenvalue or an EP of $H$. The proof is direct when the imaginary part of $\Omega$ is non-zero since
\begin{equation}
    |\Omega|^2 \kappa(X) = \bar{\Omega} \bar{X}^\top UHX = \bar{\Omega} \bar{X}^\top \bar{H}^\top UX = \bar{\Omega}^2 \kappa(X) \ .
\end{equation}

When $\Omega$ is strictly real (non-zero) but is an EP of $H$, there is an eigenvector $X_0$ of $H$ for this eigenvalue that belongs to a Jordan chain, i.e. there is an associated vector $X_1$ such that $HX_1 =\Omega X_1 + X_0$. The property $\kappa (X_0) = 0$ then follows from the equality
\begin{equation}
    \begin{split}
    &(\bar{H} \bar{X_0})^\top U HX_1 = \bar{\Omega} \bar{X_0}^\top U X_0 + |\Omega|^2 \bar{X_0}^\top U X_1 \\
    &=\bar{X_0}^\top U H^2 X_1 = 2 \Omega \bar{X_0}^\top U X_0 + \Omega^2 \bar{X_0}^\top U X_1 \ .
    \end{split}
\end{equation}
When $\Omega$ is real, non-degenerated or simply degenerated (i.e. the algebraic and geometric multiplicities of $\Omega$ are equal), $\kappa \ne 0$ \cite{mackay2020stability}.

Fig.~\ref{fig:krein} shows the Krein quantities of the four eigenvectors of $H$ for $L_\ell = 0$, as the parameters vary such that the EP ring is crossed. Outside of the ring, the symbol $H$ is diagonalizable and all its eigenvalues are real. As such, $\kappa \ne 0$. At the EP ring, three of the four eigenvectors of $H$ eventually merge into one and $H$ is no longer diagonalizable  (appendix \ref{app:EPs}). A Krein collision occurs at the EP ring between the quantities $\kappa$ of three eigenvectors. Two of the corresponding eigenvalues become unstable inside the ring. Krein quantities could have been used to directly identify the location of the collision as an EP of $H$, since at least two stable eigenvalues with non-zero Krein signatures must collide for unstable eigenvalues with zero Krein signature to appear. This result is generic of non-Hermitian operators with pseudo-Hermitian symmetry \cite{kirillov2021nonconservative,mackay2020stability}.\\

\begin{figure}[H]
    \centering
    \includegraphics[width=0.3\textwidth]{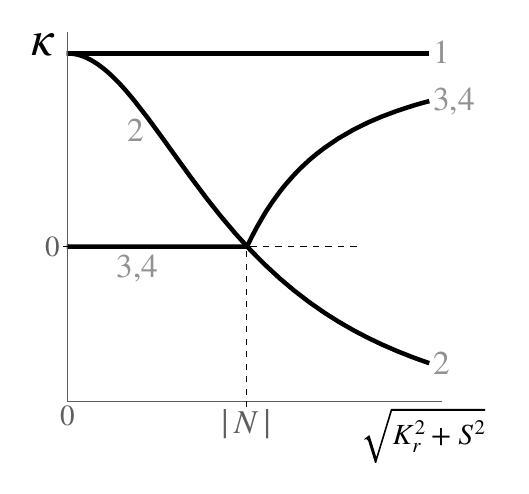}
    \caption{Krein quantities of the four eigenvectors of $H$ for $L_\ell = 0$, ordered in agreement with \eqref{eq:eigenvectorss}. At the EP ring, the three eigenvectors 2,3 and 4 are colinear and their Krein quantities vanish. Inside the EP ring ($K_r^2 + S^2 + N^2 < 0$), the Krein quantities corresponding to the unstable modes are zero.
    }
    \label{fig:krein}
\end{figure}

Let us  now consider a solution $X$ of the general linearized problem 
\begin{equation}
i \partial_t X = \mathcal{H}_5 X
\label{eq:Schroding}
\end{equation}
given by the operator \eqref{eqApp:op_pb} without projecting on the spherical harmonics. $X$ does not need to be an eigenmode of $\mathcal{H}_5$. The relation
\begin{equation}
U \mathcal{H}_5 U^{-1} = \mathcal{H}_5^\dagger ,
\label{eq:sym5}
\end{equation}
still holds, provided that $N^2$ is negative on the entire domain and suitable boundary conditions, and for the canonical scalar product
\begin{equation} \label{eq:scalar_product}
    \langle X_1 , X_2 \rangle = \int dr d\Phi \ \bar{X_1}^\top X_2 \ ,
\end{equation}
where ${\rm d} \Phi = \sin \theta {\rm d} \phi {\rm d} \theta$ is the solid angle. Note the absence of $r^2$ in the definition of the scalar product \eqref{eq:scalar_product}, owing to the change of variables \eqref{eq:transform}. Equations.~\eqref{eq:Schroding}-- \eqref{eq:sym5} ensure that the perturbation of the Hamiltonian,
\begin{equation} \label{eq:csttobs}
    \langle X , U X \rangle = \int dr d\Phi \left( |\Tilde{v}|^2 + |\Tilde{w}|^2 + |\Tilde{p}|^2 - |\Tilde{\Theta}|^2 \right) ,
\end{equation}
is constant in time. When $X$ is an eigenmode of $\mathcal{H}_5$, Eq.~\ref{eq:csttobs} is proportional to the Krein quantity previously defined. For an unstable mode, $\langle X , U X \rangle = 0$ thus implies an opposite evolution of kinetic energy and available potential energy:
\begin{equation}
    \int dr d\Phi \left( |\Tilde{v}|^2 + |\Tilde{w}|^2 + |\Tilde{p}|^2 \right) = \int dr d\Phi |\Tilde{\Theta}|^2 \ .
\end{equation}

\section{Normal form}
\label{app:analSol}

Setting $N^2< 0$, $c_{\rm s}$ and $L_\ell = (c_\mathrm{s}^2{\ell(\ell+1)}/{r^2})^{1/2}$ to constant values, Eq.~\eqref{eq:schro} reduces to 
\begin{eqnarray}
\left(-\omega^2-\left|N^{2}\right|\right) \Tilde{w}&=& i\omega\left(c_s \partial_{r}-S\right) \Tilde{p}, \label{eq:ODE1}\\
\left(-\omega^2+L_{\ell}^{2}\right) \Tilde{p}&=&i\omega\left(c_s \partial_{r}+S\right) \Tilde{w}.\label{eq:ODE2}
\end{eqnarray} 
The simplest function form that allows $S$ to cancel in a single point is
\begin{equation}
S(r) =\pm \alpha\left(r-r_{0}\right),
\label{eq:Sanalytic}
\end{equation}
where $\alpha > 0$. Rescaling the independent variable by ${x = \sqrt{\frac{2 \alpha}{c_\mathrm{s}}}(r-r_0)}$, one obtains after some algebra 

\begin{equation}
\left(\frac{d^{2}}{d x^{2}}+(-\frac{1}{4} x^{2} \mp \frac{1}{2}+\frac{c_{s}}{2 \alpha} k^{2})\right) \Tilde{p} = 0,
\label{eq:OH}
\end{equation}
where $k^2 \equiv \frac{(-\vert N^2\vert-\omega^2)(L_\ell^2-\omega^2)}{c_\mathrm{s}^2\omega^2}$. The solution of Eq.~\eqref{eq:OH} is a Parabolic Cylinder Function $\mathcal{U}$
\begin{equation}
\Tilde{p} = p_0 \;\; \mathcal{U}\left(\pm \frac{1}{2}-\frac{c_{s}}{2 \alpha} k^{2}, \;\; x\right). 
\label{eq:Ufunc}
\end{equation}
Enforcing regularity at infinity imposes the quantization condition
\begin{equation}
    \frac{c_{s}}{\alpha} k^{2} = 2n + (1 \pm 1),
    \label{eq:quantif}
\end{equation}
for $n \in \mathds{N}$. When $S\left(r \right) = - \alpha \left(r - r_{0} \right)$, the solution $n=0$ or equivalently $k = 0$ corresponds to the mode $\omega^2 = L_\ell^2$. The eigenfunctions  satisfy
\begin{eqnarray}
\Tilde{v},\Tilde{p} \; &\propto& \exp\left(-\frac{\alpha}{2c_\mathrm{s}}(r-r_0)^2\right),\\
\Tilde{w},\Tilde{\Theta} \; &=&0.
\end{eqnarray}
This is the Lamb-like wave, and we find that it is still propagating for negative $N^2$ in this normal form. When $S\left(r \right) = + \alpha \left(r - r_{0} \right)$, there exists a non-zero solution $X = \begin{pmatrix} \Tilde{v} & \Tilde{w} & \Tilde{\Theta} & \Tilde{p} \end{pmatrix}^\top$ which has zeros on $\Tilde{p}$ and $\Tilde{v}$, but non-zero components for $\Tilde{w}$ and $\Tilde{\Theta}$. Solving Eqs.~\eqref{eq:ODE1}-\eqref{eq:ODE2} for this case, we find the eigenmode 
\begin{eqnarray}
\Tilde{v},\Tilde{p} \; &=& 0, \\
\Tilde{w},\Tilde{\Theta} \; &\propto& \exp\left(-\frac{\alpha}{2c_\mathrm{s}}(r-r_0)^2\right),
\end{eqnarray}
with an eigenvalue $\omega^2 =-{\vert N^2 \vert}$ that formally corresponds to a mode $n^* = -1$ or equivalently, $k = 0$.
Fig.\ref{fig:NHspecFlow} shows the eigenfrequencies of this problem. Topology ensures that continuous deformations of the profile of the normal form provide modes of same nature, as shown in Fig.\ref{fig:toyStars}.

\clearpage
\section{Low-mass star formation}
\label{app:secondCore}
Figure~\ref{figApp:secondCore} shows the characteristic frequencies $N^2$ and $S$ extracted from two-dimensional axisymmetric hydrodynamical simulations of gravitational collapse of a molecular cloud core \cite{bhandare2020}, at the onset of protostar formation. The protostar, located at ${r = 0.07}$au, is delimited by a discontinuity in the density profile as a result of the second accretion shock. Outside of the shock, the contribution of the term $c_s / r - c_s ' / 2$ to the stratification parameter $S$ given by \eqref{eq:defS} is negligible and the density profile is flat enough as a byproduct of the first accretion shock. As such
\begin{equation}
    \frac{N^2}{\vert N\vert} = -\frac{g}{c_\mathrm{s}} \simeq S.
\end{equation}
Although the density profile is not a steady state \cite{vaytet2013}, the dynamical time is longer than $\sim \left|N\right|^{-1}$, the time over which convection develops, allowing for predictions from perturbations linear analysis. 
\end{widetext}

\begin{figure*}
    \centering
    \includegraphics[width=\textwidth]{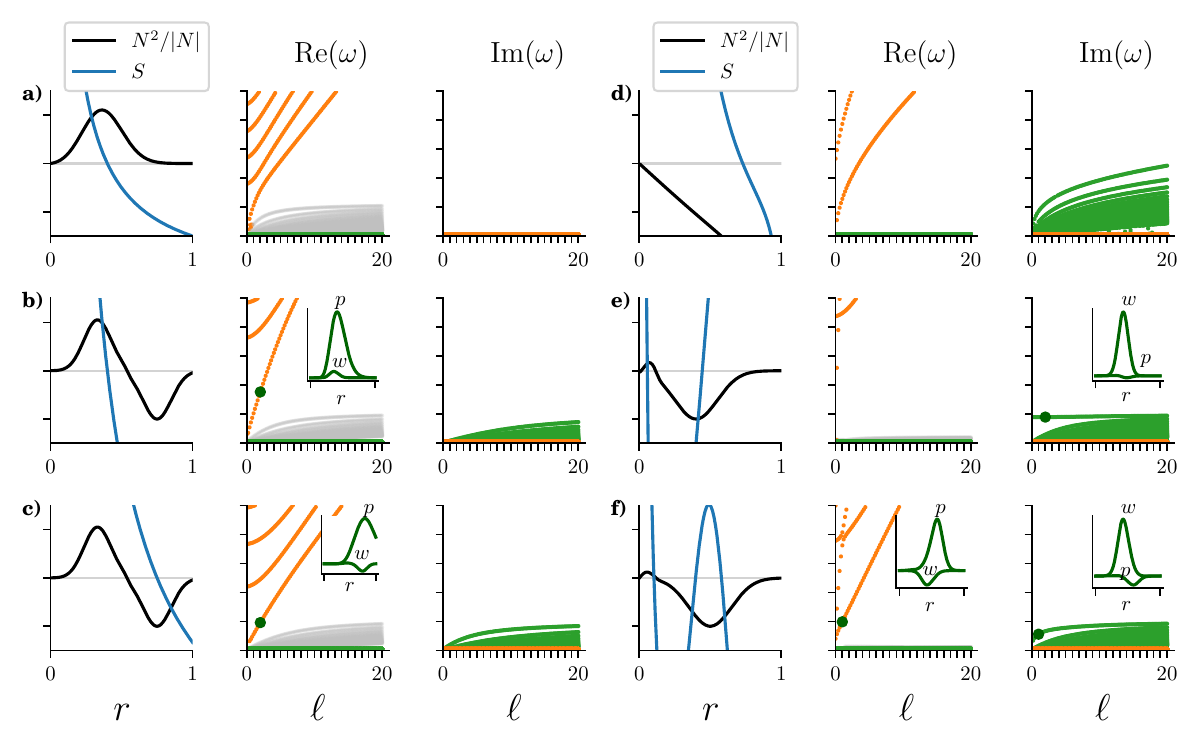}
    \caption{Complex eigenfrequencies for different profiles of unstable equilibria, obtained after numerical integration with \texttt{Dedalus}. Orange dots correspond to acoustic modes and Lamb-like wave, gray dots correspond to g-modes, green dots correspond to unstable buoyancy modes. \textbf{a)} Reference case: stable stratification discussed by \cite{leclerc2022}, where p-modes, g-modes and a Lamb-like wave propagate. \textbf{b)} A stably stratified core lies under a pre-convective outer layer. The Lamb-like wave propagates mainly in the core. \textbf{c)} Same as b), with a Lamb-like wave propagating in the unstable layer. \textbf{d)} In a $n=1$ polytropic star, $N^2$ is negative everywhere, and $S$ monotonically decreases from $+\infty$ to $-\infty$. A Lamb-like wave propagates among unstable buoyancy modes. \textbf{e)} In a star with increasing $S$ in an unstable layer, a radial unstable mode exists. \textbf{f)}  In an unstable region, each cancellation of $S$ leads to a mode described by the analytical solutions that depend on the sign change, with possible hybridization. The properties of all models are consistent with the predictions derived analytically with the normal form.}
    \label{fig:toyStars}
\end{figure*}

\begin{figure*}
    \centering
    \includegraphics[width=0.3\textwidth]{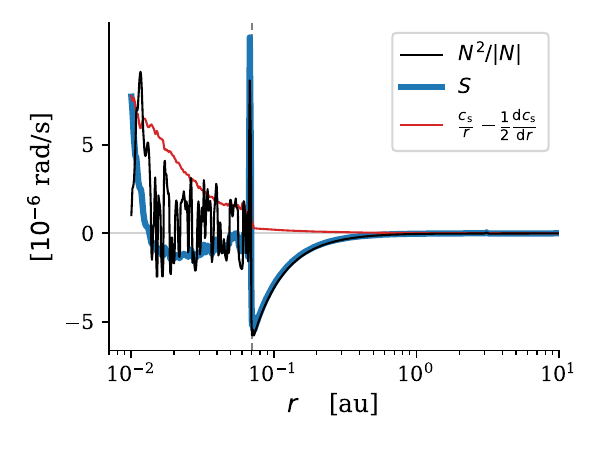}
    \includegraphics[width=0.3\textwidth]{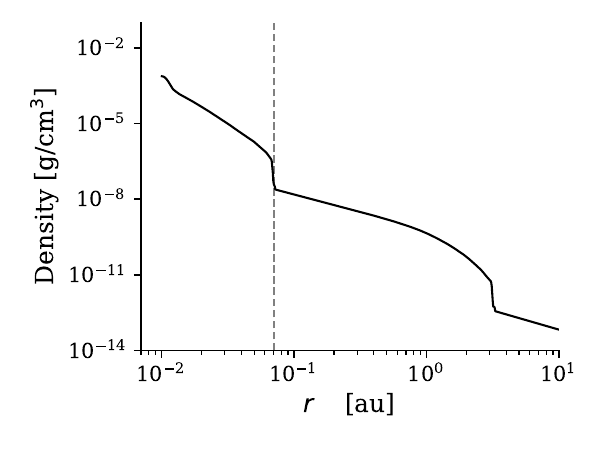}
    \includegraphics[width=0.3\textwidth]{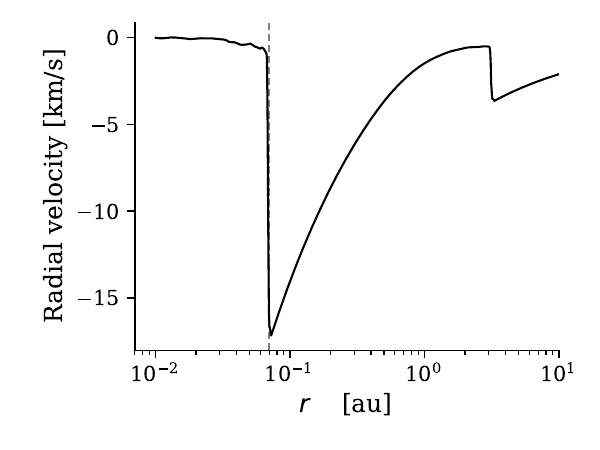}
    \caption{Profiles of different quantities from the simulation of gravitational collapse of an astrophysical cloud resulting in the formation of a low-mass protostar with an initial size of $R_1$ = 0.07 au. Outside the core, $S$ and $N^2/|N|$ are approximately equal.}
    \label{figApp:secondCore}
\end{figure*}

\end{document}